\documentclass[prl,twocolumn,superscriptaddress,amsmath,amsfonts,
               floatfix]{revtex4}

\usepackage{epsfig,psfrag}
\usepackage{CJK}

\usepackage{graphicx}
\usepackage{dcolumn}
\usepackage{bm}
\usepackage{amssymb}
\usepackage{natbib}
\usepackage{upgreek}

\begin{document}
\title{Microscopic theory of Raman scattering for the rotational organic cation in metal halide perovskites}
\author{Yu Cui}
\affiliation{Tianjin Key Laboratory of Low Dimensional Materials Physics and Preparing Technology, Department of Physics, School of Science, Tianjin University, Tianjin 300354 China}
\author{Yi-Yan Liu}
\affiliation{Tianjin Key Laboratory of Low Dimensional Materials Physics and Preparing Technology, Department of Physics, School of Science, Tianjin University, Tianjin 300354 China}
\author{Jia-Pei Deng}
\affiliation{Tianjin Key Laboratory of Low Dimensional Materials Physics and Preparing Technology, Department of Physics, School of Science, Tianjin University, Tianjin 300354 China}
\author{Xiao-Zhe Zhang}
\affiliation{Tianjin Key Laboratory of Low Dimensional Materials Physics and Preparing Technology, Department of Physics, School of Science, Tianjin University, Tianjin 300354 China}
\author{Ran-Bo Yang}
\affiliation{Tianjin Key Laboratory of Low Dimensional Materials Physics and Preparing Technology, Department of Physics, School of Science, Tianjin University, Tianjin 300354 China}
\author{Zhi-Qing Li}
\affiliation{Tianjin Key Laboratory of Low Dimensional Materials Physics and Preparing Technology, Department of Physics, School of Science, Tianjin University, Tianjin 300354 China}
\author{Zi-Wu Wang}
\email{wangziwu@tju.edu.cn}
\affiliation{Tianjin Key Laboratory of Low Dimensional Materials Physics and Preparing Technology, Department of Physics, School of Science, Tianjin University, Tianjin 300354 China}

\begin{abstract}
A gap exists in microscopic understanding the dynamic properties of the rotational organic cation (ROC) in the inorganic framework of the metal halide perovskites (MHP) to date. Herein, we develop a microscopic theory of Raman scattering for the ROC in MHP based on the angular momentum of a ROC exchanging with that of  the photon and phonon. We systematically present the selection rules for the angular momentum transfer among three lowest rotational levels. We find that the phonon angular momentum that arising from the inorganic framework and its specific values could be directly manifested by Stokes (or anti-Stokes) shift. Moreover, the initial orientation of the ROC and its preferentially rotational directions could be judged in Raman spectra. This study lays the theoretical foundation for the high-precision resolution and manipulation of molecular rotation immersed in many-body environment by Raman technique.
\end{abstract}
\maketitle

Over the past few years, metal halide perovskites (MHP) as promising materials have aroused the intense interest from the worldwide research owing to their notable properties in the fields of photovoltaic cells, light emitting diodes, and photodetectors\cite{w1,w2,w3}. Traditionally, the consensus is that the species of organic cations are not directly involved in the formation of electronic transport levels\cite{w4,w5}. However, the recent breakthroughs, studied by the spectral measurements\cite{w6,w7,w8} and the first-principles calculation\cite{w12,w13,w14,b1}, have shown that the dipole nature of the organic cation plays a critical role in the structure stabilizations and optoelectronic properties of MHP\cite{w15}, e.g. the compositional engineering of organic cations to modulate the bandgap and to modify the crystal symmetry and phase. In particular, the rotational motion of organic cation\cite{b1,w15,w9,w10,w11} results in an effective Coulomb screening to affect the dynamic of charge carriers. Therefore, the manipulation of the rotational organic cation (ROC) not only gives an effective method to modify the properties of MHP, but provides a test bed to explore novel quantum phenomena in many-body physics\cite{w15,w17}.

Based on angular momentum exchange between photon and rotational particles, quantum control of the rotational atoms or molecules by laser have been studied both theoretically and experimentally in areas of atomic, molecular and optical physics as well as in physical chemistry\cite{w16,w17,w18,w19}. While the corresponding studies on ROC in perovskite materials are still very few. On the other hand, ROC inevitably couples with the surrounding inorganic cage [see Fig. 1(a)]. Although the coupling effect of organic cation with inorganic sublattice by the hydrogen bonds have been analyzed widely\cite{w53,w55}, the role of phonons of inorganic cage on the rotational dynamic of organic cation received relatively little attention, partially because of the intricate angular-momentum algebra from the ROC coupled with many-body environment. Fortunately, Schmidt and Lemeshko undertook a critical step towards such a theory in 2015 by introducing the quasiparticle concept of the ``$angulon$"--a quantum rotor dressed by a bath of harmonic oscillators\cite{w30}, which provides a simple and effective model to study the angular momentum exchange between the rotational molecule and many-body environment, such as molecules rotating in superfluid helium, ultracold alkali dimers interacting with a Bose-Einstein condensate\cite{w20,w21}. However, to the best of our knowledge, the theoretical model for the angular momentum of ROC exchanges with both photon and phonon in MHP has not yet been developed.

\begin{figure}
\includegraphics[width=3.3in,keepaspectratio]{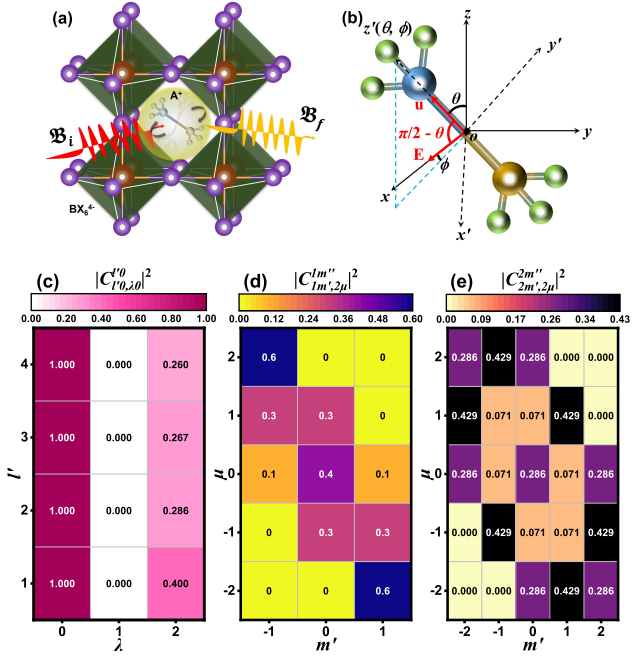}
\caption{\label{compare} (a) The schematic diagram of Raman scattering for the rotational A$^+$ cation in the centre of BX$_6^{4-}$ octahedral cage, where A, B, and X correspond to the species of organic cation, metal ion, and halide anion, respectively, in MHP. $ {\mathfrak{B} _i}$ and ${\mathfrak{B} _f}$ represent the angular momenta (in the unit of rotational constant $B$) of the incident and scattering photon, respectively. (b) The direction of the organic cation in the laboratory-frame ($x$, $y$, $z$) and molecular-frame ($x'$, $y'$, $z'$). ($\theta $, $\phi$) denotes the angular coordinates of organic cation in the laboratory frame. $\bf{u }$ is the molecular dipole moment and ${\bf{E}}$ is the electric field vector of the light. (c-e) Clebsch-Gordan coefficients for different orbital quantum states, $\left| {L=l',M=m'} \right\rangle$, coupling with phonon angular momentum states, $\left| {\lambda ,\mu } \right\rangle$.}
\end{figure}

In this paper, we study the microscopic processes of Raman scattering mediated by a ROC in MHP based on ROC coupling with photon and phonon. We present the selection rules of quantum transitions among the rotational eigenstates $\left| {L,M} \right\rangle $ of ROC ($L$ and $M$ denote the orbital angular quantum number and its projection on the laboratory-frame $z$-axis, respectively.), in which transitions from $L=l$ to $L=l'$ accompanying with and without the variation of the projection of angular quantum number are analyzed. We illustrate the Stokes and anti-Stokes shifting of Raman spectra for three lowest rotational levels according to different angular momenta provided by phonons. These results show that the transfer of phonon angular momentum, the initial orientation and the magnitude of the rotational angle of a ROC could be reflected by Raman scattering, both of which are the key problems for accurately quantum control of the ROC or molecules. More importantly, this model can be expanded to study molecules rotating in varieties of the cage-like structures, such as fullerene, carbon nanotube and so on.

In the frame of the classical model of Raman scattering mediated by elementary excitation, e.g. electron and exciton\cite{w22,w23,w24}, Raman scattering for ROC in MHP could be divided into three steps as schemed in Fig. 1(a): (i) the angular momentum transfer of an incident photon ${\mathfrak{B} _i}$ excites the ROC from the initial state $\left| i \right\rangle  =\left| {L=l,M=m} \right\rangle$ to $\left| j \right\rangle  =\left| {L=l',M=m'}\right\rangle$, where these rotational eigenstates are labeled by the orbital angular number $L$ and its projection, $M$, on the laboratory-frame $z$-axis, with eigenenergies ${E_L} = BL(L+ 1)$\cite{w25,w26}, $B$ is the rotational constant; (ii) the transition from $\left| j \right\rangle  = \left| {L=l',M=m'} \right\rangle$ to $\left| k \right\rangle  = \left| {L=l'',M=m''} \right\rangle$ is accompanied by the exchange of angular momentum between ROC and phonons; (iii) ROC from $\left| k \right\rangle$ comes back into the initial $\left| i \right\rangle$ with the help of the angular momentum transfer $ {\mathfrak{B} _f}$ by the scattering of a photon.
So the cross section of Raman scattering for ROC can be expressed as\cite{w22,w23,w24}
\begin{equation}
{\left| \Re  \right|^2} = {\left| {\frac{{\sum\nolimits_{jk} {\left\langle {i\left| {{{\hat H}_{opt}}} \right|j} \right\rangle \left\langle {j\left| {{{\hat H}_{ph}}} \right|k} \right\rangle \left\langle {k\left| {{{\hat H}_{opt}}} \right|i} \right\rangle } }}{{\left[ { {\mathfrak{B} _i} - \left( {{E_j} - {E_i}} \right) - i\Gamma } \right]\left[ { {\mathfrak{B} _i} -  {\mathfrak{B} _f} \pm {E_0} - i\Gamma } \right]}}} \right|^2},\\
\end{equation}
where $E_i$ ($E_j$) is the eigenenergy of rotational state $\left| i \right\rangle$ ($\left| j \right\rangle$), $E_0$ is the rotational energy provided by phonons in unit of $B$, and $\Gamma$ is the homogeneous line-width for the quantum transition. ${\langle {i| {{{\hat H}_{opt}}} |j} \rangle }$ (${\langle {k |{{{\hat H}_{opt}}} |i} \rangle }$) and ${\langle {j| {{{\hat H}_{ph}}} |k} \rangle }$ are the matrix elements for transitions between different rotational states, arising from ROC-photon and -phonon interaction, respectively. They are the key components to study the Raman scattering of ROC in the following.

For the sake of clarity, the ROC, such as CH$_3$NH$_3^+$ in MHP, is regarded as a linear molecule with frozen transitional motion. In the dipole approximation, the Hamiltonian for a ROC subjected to a linearly polarized light is given by ${\hat H_{opt}} =  -{{\bf{u }} \cdot {\bf{E}}}$, where $\mathbf{u}$ is the inherent dipole moment for ROC oriented along $z'$-axis in the molecular-frame and ${\bf{E}}$ is the electric field vector of the light along the $x$-axis in the laboratory-frame\cite{w27,w28,w29}, while the relative orientation of these two frames is given by the Euler angle ($\phi$, $\theta$) as shown in Fig. 1(b).  The ROC-light interaction, in general, regarded as the perturbation to the system, gives rise to quantum transitions between different rotational states. After a series of mathematical processes (see the supplemental materials), the corresponding matrix element between $\left| i \right\rangle $ and $\left| j \right\rangle $ states is expressed as

\begin{eqnarray}
\left\langle {i\left| {{{\hat H}_{opt}}} \right|j} \right\rangle&=&F(l,l',m,m')\nonumber\\
 &=&-{u}{E}{a_{l' - 1,m'}}{\delta _{l,l' - 1}}{\delta _{m,m'}}\nonumber\\
  &&  + {u}{E}{b_{l' - 1,m' - 1}}{\delta _{l,l' - 1}}{\delta _{m,m' - 1}}\nonumber\\
  &&- {u}{E}{b_{l' - 1, - (m' + 1)}}{\delta _{l,l' - 1}}{\delta _{m,m' + 1}},
\end{eqnarray}
where ${a_{l,m}}$ and ${b_{l,m}}$ are angular momentum dependent coefficients, shown in the supplemental materials. The constants ${u }$ and $E$ represent the magnitude of ${\bf{u }}$ and ${\bf{E}}$. From Eq. (2), it can be inferred that the angular momentum of the incident photon results in the transition between orbital quantum states follows the selection rule of $l'-l=1$ along with the exchange of angular momentum projection $m'-m=0,\pm1$ in the step (i) of Raman scattering. Similarly, the emission of photon in the step (iii) satisfies the selection rule of $l-l''=1$ along with $m-m''=0,\pm1$.

In the frame of $angulon$ model, the effective Hamiltonian describing the interaction between a ROC and phonon bath in the spherical basis is given by\cite{w30,w31,w32,w33}:
\begin{eqnarray}
\hat H_{ph} &=& \sum\limits_{q\lambda \mu } {{V_\lambda }(q)\left[ {\hat Y_{\lambda \mu }^*(\hat \theta ,\hat \phi )\hat b_{q\lambda \mu }^\dag  + {\hat Y_{\lambda \mu }}(\hat \theta ,\hat \phi ){{\hat b}_{q\lambda \mu }}} \right]}.
\end{eqnarray}
Here $q = \left| {\bf{q}} \right|$ is the scalar representation of the phonon wave vector, satisfying the relation $\sum\nolimits_q  \equiv  \int {dq}$. $\lambda$ and $\mu$ define, respectively, the phonon angular momentum and its projection onto the $z$-axis. $\hat Y_{\lambda \mu }(\hat \theta ,\hat \phi )$ are the spherical harmonic operators, which is essential for the microscopic description of the transfer of phonon angular momentum. ${\hat b_{q\lambda \mu }^\dag }$ and ${\hat b_{q\lambda \mu } }$ are the creation and annihilation operators of phonons in the angular momentum representation, respectively (see Refs. \cite{w30} and \cite{w31} for the detailed derivation). The angular momentum dependent interaction potential ${V_\lambda }{\rm{ }}(q) = {v_\lambda }{\left[ {8{\alpha _c}{q^2}/\left( {2\lambda  + 1} \right)} \right]^{1/2}}\int {dr{r^2}{f_\lambda }(r){j_\lambda }(qr)}$ is employed for an organic cation rotating in the cage-like phonon bath, where $\alpha_c$ is the Fr$\ddot{o}$hlich coupling constant, meaning that the exchange of angular momentum between longitudinal optical (LO) phonons and ROC are mainly taken into account\cite{w31,w32,w33}, because LO phonon is the dominate mode caused by the lattice distortion of the cage-like structure both in the theoretical and experimental perspectives\cite{w34,w35,w36,w37}; ${{j_\lambda }(qr)}$ is the spherical Bessel function; ${v_\lambda }$ and ${{f_{\lambda} }(r)}$ represent the strength and the shape of the potential in the respective angular momentum channel $\lambda$. Especially, the latter function describes the microscopic details of two-body interaction between ROC and phonon bath, whose expression is proposed as\cite{w40,w41}
\begin{equation}
{{ {{{{f_{\lambda} }(r)}}} } } = \left\{\begin{array}{rcl}
&{(\frac{r}{R})^{\lambda} },& (r \le R)\\
&0,& (r > R)\\
\end{array}\right.,
\end{equation}
and the octahedral inorganic cage is approximated by the spherical cavity, based on the facts that (i) the rotating behavior of the organic cation in this cage demonstrated widely by recent experiments\cite{w10,w11,w38,w39}; (ii) the strongest coupling strength (the potential distribution) is around the spherical boundary between ROC and inorganic cage\cite{ww1,ww2}. $R=a_0/2$ is the effective radius of this spherical space ($a_0$ is the length of the side of the octahedral cage). After proceeding some algebraic calculation, the matrix element for the transfer of phonon angular momentum between two different rotational states is given as
\begin{eqnarray}
\left\langle {j\left| {{{\hat H}_{ph}}} \right|k} \right\rangle
= \sum\limits_q{V_\lambda }(q)g_1 C_{l'0,\lambda 0}^{l''0}C_{l'm',\lambda \mu }^{l''m''},
\end{eqnarray}
where $g_1={\left\{ {\left( {2l'{\rm{ + 1}}} \right)(2\lambda  + 1)/\left[ {4\pi \left( {2l'' + 1} \right)} \right]} \right\}^{1/2}}$, and $C_{l'm',\lambda \mu }^{l''m''}$ is the Clebsch-Gordan (C-G) coefficients\cite{w42}. Eventually, upon substitution of Eqs. (2) and (5), the cross section of Raman scattering is converted into
\begin{eqnarray}
{\left| \Re  \right|^2} = {\left| {\frac{{u^2E^2{g_1}{g_2}\sum\nolimits_q {{V_\lambda }\left( q \right)} C_{l'0,\lambda 0}^{l''0}C_{l'm',\lambda \mu }^{l''m''}}}{{\left[ {{{\left( {{\mathfrak{B} _i} - {\mathfrak{B} _f} \pm {E_0}} \right)}^2} + {\Gamma ^2}} \right]}}} \right|^2}.
\end{eqnarray}
${g_2} = F(l,l',m,m')F(l'',l,m'',m)$ summaries the roles of angular momentum transfer between ROC and photon in the absorption and emission processes, ensuring the classical process of Raman scattering; namely, the initial and final states are the same one. The redistribution of angular momentum between ROC and phonon bath in mediated process is determined by the C-G coefficients $C_{l'0,\lambda 0}^{l''0}$ and $C_{l'm',\lambda \mu }^{l''m''}$.

The values of $|C_{l'0,\lambda 0 }^{l''0}{|^2}$ as functions of $l'$ and $\lambda$ are shown in Fig. 1(c), in which $l''=l'$ is assumed, implying the orbital angular quantum is unchanged during the angular momentum exchange between ROC and phonon. Thus, the phonon angular momentum only induces the change of the projection of orbital angular momentum, that is the variation of the orientation of ROC. This results in the distribution of the allowed transitions (red) as well as forbidden ones (white) depending on the phonon angular-momentum of quantum state $\lambda$, satisfying the following selection rule of $l' + l' + \lambda=$ even. As a result, the lowest order of phonon angular-momentum that dominates the transfer of angular momentum between ROC and phonon is the quantum number $\lambda=2$ shown in Fig. 1(c).  Figs. 1(d) and (e) present the phonon angular-momentum state $\lambda=2$ induces possible transition between the projections of angular momentum reflected by C-G coefficients $|C_{1m',\lambda\mu}^{1m''}{|^2}$ and $|C_{2m',\lambda\mu}^{2m''}{|^2}$ for angular quantum number $l'=1$ and $l'=2$, respectively. Two obvious features are shown that (1) the distribution of the coefficients reveals the symmetrical relations, represented as $|C_{l'm',\lambda \mu }^{l''m''}{|^2} = |C_{l' - m',\lambda  - \mu }^{l'' - m''}{|^2}$; (2) the ROC is inclined to couple with phonons that can invert its angular momentum projection from $m'$ to $-m'$. These results indicate that ROC has the preferential directions induced by phonon angular momentum.

In order to give the comprehensive comparison between different transfer of phonon angular momentum, the Raman spectra for three lowest rotational levels are illustrated in Fig. 2. Here, the typical example of MHP CH$_3$NH$_3$PbI$_3$ in cubic phase is selected, in which organic cation CH$_3$NH$^+_3$ rotating in the PbI$_6^{4-}$ octahedral cage as schemed in Fig. 1(a). The values for the related parameters in Eq. 6 are listed in Table S1 in the supplemental materials. These specific values of angular momentum dependent coefficients $a_{l,m}$, $b_{l,m}$ and C-G coefficients involved in the angular momenta transfer among three lowest rotational levels are listed in Table S3 and Table S4 in the supplemental materials.

\begin{figure*}[t!]
\centering
\includegraphics[width=7in,keepaspectratio]{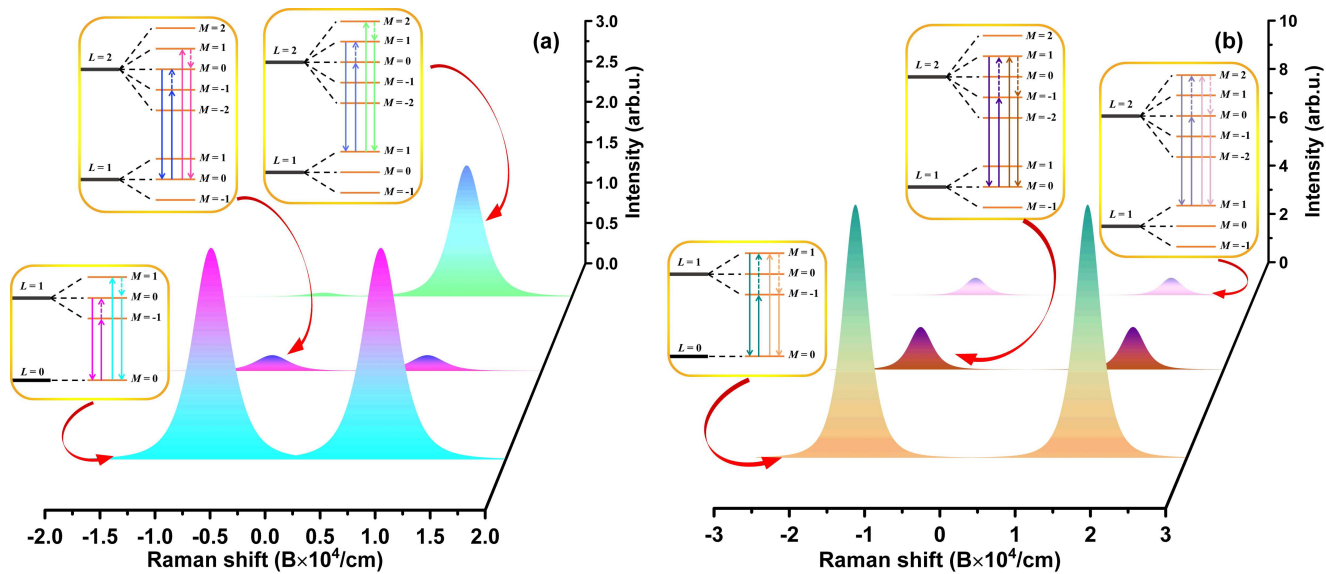}
\caption{The Stokes and anti-Stokes Raman scattering for three lowest rotational levels coupling with the phonon angular momentum states $\left| {\lambda=2,\mu=\pm1} \right\rangle$ (a) and $\left| {\lambda=2,\mu=\pm2} \right\rangle$ (b), respectively. The senarios of the angular momentum transfer for different Raman processes are shown in the insets, where $L$ and $M$ represent the quantum number of angular momentum and its projection onto the $z$-axis, respectively.}
\end{figure*}

Firstly, Fig. 2(a) illustrates the Raman scattering from the ground state of ROC ($L=0$) to the first-excited state ($L=1$) as well as from $L=1$ to the second-excited state ($L=2$) with the change of projection of angular momentum $\triangle{M}=\pm1$. One can see that the Stokes process of  ($L=0,M=0$)$\rightarrow$($L=1,M=1$)$\dashrightarrow$($L=1,M=0$)$\rightarrow$($L=0,M=0$) [$\rightarrow$ and $\dashrightarrow$ denote the transfer of angular momentum of photon and phonon, respectively.], and anti-Stokes process of ($L=0,M=0$)$\rightarrow$($L=1,M=-1$)$\dashrightarrow$($L=1,M=0$)$\rightarrow$($L=0,M=0$) follow a symmetrical distribution with the same intensity, which can be attributed to the symmetrical relations of C-G coefficient, shown in Fig. 1(d), and the optical coefficients given in Eq. (2). In 2015, Schmidt and Lemeshko proposed the phonon angular momentum couples with the rotating quantum molecule (or impurity) to form a new quasiparticle--$angulon$ for the first time\cite{w30}. They pointed out this angulon induces a rich rotational fine structure in spectra of molecules, such as ``rotational Lamb shift"; subsequently, they further revealed that the spectral function of the rotational molecule suddenly acquires the transfer of one quantum of phonon angular momentum from the many-body environment at a critical rotational speed; however, the direct identification for the phonon angular momentum and its transfer is still a challenge task in experiments. For Raman scattering in Fig. 2, not only the phonon angular momentum arising from the octahedral-cage structure is proved, but also its specific values could be reflected directly by the Stokes and anti-Stokes shifts. In Fig. 2(a), the Stokes process of ($L=1,M=0$)$\rightarrow$($L=2,M=1$)$\dashrightarrow$($L=2,M=0$)$\rightarrow$($L=1,M=0$) and the anti-Stokes process of ($L=1,M=0$)$\rightarrow$($L=2,M=-1$)$\dashrightarrow$($L=2,M=0$)$\rightarrow$($L=1,M=0$) show the similar behaviors, however, with the smaller intensity, since the different values between $C_{1m',2\mu }^{1m''}$ and $C_{2m',2\mu }^{2m''}$ given in Figs. 1(d) and (e), respectively. But the intensity of the Stokes process ($L=1,M=1$)$\rightarrow$($L=2,M=2$)$\dashrightarrow$($L=2,M=1$)$\rightarrow$($L=1,M=1$) is much stronger than that of the anti-Stokes process ($L=1,M=1$)$\rightarrow$($L=2,M=0$)$\dashrightarrow$($L=2,M=1$)$\rightarrow$($L=1,M=1$). Following the rules of C-G coefficients in Fig. 1(e), the asymmetrical intensity distribution should also be appeared for these scattering processes starting from the initial states ($L=1,M=-1$) (see Fig. S2 in the supplemental materials). From these comparison, we can infer that the difference of the initial states between ($L=1,M=0$) and ($L=1,M=\pm1$), that is, the differently initial orientation of ROC, determines the features of Raman spectra. In turn, the initial orientation of ROC (or molecules) could be reflected by Raman spectra in experiments. In fact, a series of strategies to control the alignment and orientation of the rotational molecules have been proposed in the past decades, such as the linearly polarized ultrafast laser pules\cite{wy1,wy2,wy3}, two-color and static fields\cite{wy4}, as well as two-color femtosecond lasers and terahertz field\cite{wy6,wy7}.
The molecular alignment and orientation are of crucial for a variety of applications ranging from chemical reaction dynamics to the design of molecular devices. For these applications, however, one of the most important prerequisites is to judge the initial orientation of the rotational molecules. Obviously, on the one hand, the Raman scattering of rotational molecule, provides an effective method to overcome this issue; on the other hand, the detailed dynamics and some novel physical phenomena related to the rotational particles in many-body bath should be analyzed deeply by Raman scattering even though the rotational structure and dynamics of molecules have been obtained widely from infrared spectroscopy\cite{wb1,wb2}.

Fig. 2(b) shows Raman scattering starting from the ground and first-excited states of ROC with the change of projection of angular momentum $\Delta M=\pm2$. One can see that the Stokes and anti-Stokes scattering starting from the initial states ($L=0,M=0$) and ($L=1,M=0$) have the same intensity. Moreover, the magnitude is much stronger than the corresponding scattering shown in Fig. 2(a). This indicates the rotation of organic cation has the preferentially orientation at certain external condition, which is very consistent with the prediction of C-G coefficient in Fig. 1(e). Therefore, this kind of Raman scattering would provide the anticipation for these chemical reactions depending on the high-precision control of the spatial orientation of the molecules, e.g. molecular imaging and selectivity\cite{wy11}, the enhancement of the interaction of a molecule with a surface in precise catalytic process\cite{wy12,wy13}. For Raman scattering starting from the same states ($L=1,M=\pm1$) of ROC in Figs. 2(a) and (b), the spectral shapes show the significant difference since the change of projection of angular momentum follows $\Delta M=\pm1$ and $\Delta M=\pm2$, respectively. This means that the variational magnitude of the orientation of a ROC could be estimated by the spectral shape, which suggests the possibility to judge and modulate a molecule from a well-defined initial state to a target state.

The phonon angular momentum not only induces the variation of orientation of ROC at a given orbital quantum state, but induces transitions between different orbital states in Raman scattering when the second-order term of the photon interacting with ROC is considered. We illustrate the processes of ($L=0,M=0$)$\rightarrow$($L=2,M=0$)$\dashrightarrow$($L=1,M=0$)$\rightarrow$($L=0,M=0$) and ($L=0,M=0$)$\rightarrow$($L=1,M=0$)$\dashrightarrow$($L=2,M=0$)$\rightarrow$($L=0,M=0$) without the variation of orientation as well as the processes of ($L=0,M=0$)$\rightarrow$($L=2,M=0$)$\dashrightarrow$($L=1,M=-1$)$\rightarrow$($L=0,M=0$) and ($L=0,M=0$)$\rightarrow$($L=1,M=-1$)$\dashrightarrow$($L=2,M=0$)$\rightarrow$($L=0,M=0$) with the change of orientation $\Delta M=\pm1$ in Fig. S3 (see the supplemental materials). One can see that these scatterings have the similar features with the processes in Fig. 2, but the intensity become weaker by nearly one order of magnitude, demanding more accurate detection techniques. In fact, these scattering behaviors for three lowest rotational levels can be generalized to more quantum transitions between rotational levels assisted by photon and phonon angular momenta for ROC in MHP and could be explored by Raman scattering. Therefore, this type of Raman scattering could get deep into more complicated quantum transitions and reveal the fine spectroscopy of the rotational systems. Besides the rotational molecules or impurities as intermediaries for the Raman scattering, other elementary excitations in physics, such as the interlayer and intralayer excitons in van der Waals heterostructures\cite{wy14,wy15}, should have the similar Raman scattering when the rotational degree of freedom is considered. Lastly, we must emphasize that (1) only the angular momentum of LO phonon is considered in this study, other phonon modes have the similar effect and will play the important role at certain condition; (2) the perovskite materials undergo two structure phase transitions with temperature, the influence of which on the coupling strength between phonon bath and ROC, as well as the shape of the potential are also significant\cite{w3.8}. These effects are out of the scope of this study.

 In summary, we develop a microscopic theory to describe the Raman scattering of an organic cation rotating in the octahedral cage of MHP. This theory predicts that the Raman spectra provide an effective and direct method to reflect the transfer of phonon angular momentum and its specific values in many-body environment. Meanwhile, two key prerequisites for the alignment and orientation of the rotational molecules could be judged by Raman spectra, which may open a new door to explore quantum control of particle rotation in many-body physics.

This work was supported by National Natural Science Foundation of China (Grant Nos. 11674241 and 12174283).

\end{document}